\documentclass[journal=jacsat,manuscript=article]{achemso}

\usepackage[version=3]{mhchem} % Formula subscripts using \ce{}
\usepackage{subcaption}
\usepackage[nolist]{acronym}

\begin{acronym}[MPC]
\acro{DFT}{density functional theory}
\acro{WFT}{wavefunction theory}
\acro{PBC}{periodic boundary conditions}
\acro{FCI}{full configuration interaction}
\acro{CC}{coupled cluster}
\acro{CAS}{complete active space}
\acro{SA}{state-average}
\acro{fUCC}{factorized Unitary Coupled Cluster}
\acro{CASSCF}{complete active space self-consistent-field}
\acro{CASCI}{complete active space configuration interaction}
\acro{AIMP}{ab-initio model potential}
\acro{SI}{supporting information}
\acro{MO}{molecular orbital}
\acro{MAD}{mean absolute deviation}
\acro{CSF}{configuration state function}
\acro{VQE}{variational quantum eigensolver}
\end{acronym}

\author{Ernst Dennis Lægteskov Binau Larsson}
\email{edl@sdu.dk}
\affiliation[SDU]{Department of Physics, Chemistry and Pharmacy, University of Southern Denmark, Campusvej~55, DK--5230 Odense M, Denmark}
\author{Erik Kjellgren}
\affiliation[SDU]{Department of Physics, Chemistry and Pharmacy, University of Southern Denmark, Campusvej~55, DK--5230 Odense M, Denmark}
\author{Peter Reinholdt}
\affiliation[SDU]{Department of Physics, Chemistry and Pharmacy, University of Southern Denmark, Campusvej~55, DK--5230 Odense M, Denmark}
\author{Jacob Kongsted}
\affiliation[SDU]{Department of Physics, Chemistry and Pharmacy, University of Southern Denmark, Campusvej~55, DK--5230 Odense M, Denmark}

\title[]
  {State-Averaged Quantum Algorithms for Multiconfigurational Surface Chemistry: A Benchmark on Rh@TiO$_{2}$(110)}

\abbreviations{IR,NMR,UV}

\begin{document}

%%%%%%%%%%%%%%%%%%%%%%%%%%%%%%%%%%%%%%%%%%%%%%%%%%%%%%%%%%%%%%%%%%%%%
%% The "tocentry" environment can be used to create an entry for the
%% graphical table of contents. It is given here as some journals
%% require that it is printed as part of the abstract page. It will
%% be automatically moved as appropriate.
%%%%%%%%%%%%%%%%%%%%%%%%%%%%%%%%%%%%%%%%%%%%%%%%%%%%%%%%%%%%%%%%%%%%%
%\begin{tocentry}

%Some journals require a graphical entry for the Table of Contents.
%This should be laid out ``print ready'' so that the sizing of the
%text is correct.

%Inside the \texttt{tocentry} environment, the font used is Helvetica
%8\,pt, as required by \emph{Journal of the American Chemical
%Society}.

%The surrounding frame is 9\,cm by 3.5\,cm, which is the maximum
%permitted for  \emph{Journal of the American Chemical Society}
%graphical table of content entries. The box will not resize if the
%content is too big: instead it will overflow the edge of the box.

%This box and the associated title will always be printed on a
%separate page at the end of the document.

%\end{tocentry}

%%%%%%%%%%%%%%%%%%%%%%%%%%%%%%%%%%%%%%%%%%%%%%%%%%%%%%%%%%%%%%%%%%%%%
%% The abstract environment will automatically gobble the contents
%% if an abstract is not used by the target journal.
%%%%%%%%%%%%%%%%%%%%%%%%%%%%%%%%%%%%%%%%%%%%%%%%%%%%%%%%%%%%%%%%%%%%%
\begin{abstract}
Accurate modeling of surface catalytic processes often requires methods capable of describing strong correlation, charge transfer, and multiple closely lying electronic states. While density functional theory remains widely used, its limitations for localized electronic states motivate the use of wavefunction-based approaches and, more recently, quantum computing algorithms. However, the performance of quantum ansätze in chemically motivated, multistate settings remains largely unexplored.

Here, we benchmark state-averaged factorized unitary coupled cluster with singles and doubles (SA-fUCCSD) and the adaptive, problem-tailored ansatz (SA-ADAPT) using an embedded cluster model of NO adsorption on Rh-doped TiO$_{2}$(110). The system exhibits pronounced multiconfigurational character and multiple state crossings, providing a stringent test. State-averaged CASSCF serves as a reference, and the quantum ansätze are evaluated as solvers for the corresponding CASCI problem within a fixed orbital basis.

We find that SA-fUCCSD improves with increasing circuit depth but requires many parameters and shows sensitivity to initialization. In contrast, SA-ADAPT achieves near-CASSCF accuracy with significantly fewer operators. A modified operator selection scheme, incorporating multiple operators per iteration, substantially accelerates convergence.

Our results demonstrate the efficiency of adaptive ansätze for multistate problems and establish a controlled benchmark for quantum algorithms in chemically motivated systems beyond minimal models.
\end{abstract}

%%%%%%%%%%%%%%%%%%%%%%%%%%%%%%%%%%%%%%%%%%%%%%%%%%%%%%%%%%%%%%%%%%%%%
%% Start the main part of the manuscript here.
%%%%%%%%%%%%%%%%%%%%%%%%%%%%%%%%%%%%%%%%%%%%%%%%%%%%%%%%%%%%%%%%%%%%%
\section{Introduction}

Accurate modeling of surface catalytic processes remains a long-standing challenge in computational quantum chemistry. These processes typically involve materials with complex electronic structures, bond breaking and formation of adsorbed species, and significant charge transfer between surface and adsorbate. Surface defects, whether intrinsic or extrinsic, can further play non-innocent roles, complicating both theoretical descriptions and experimental interpretation.

The traditional workhorse for modeling surface catalysis is \ac{DFT} with \ac{PBC}. While PBC-DFT has proven remarkably predictive for a wide range of systems, it can struggle in situations involving localized electronic states, such as transition metal dopants. These challenges are largely attributed to the self-interaction error inherent to standard \ac{DFT} functionals~\cite{Perdew1981}. Common remedies, such as hybrid functionals or the Hubbard $+U$ correction, can partially alleviate these issues, but both introduce additional complications. Hybrid functionals are computationally demanding under \ac{PBC} and rely on an empirically chosen fraction of Hartree–Fock exchange, while the $+U$ parameter is similarly empirical and sensitive to both the targeted property and the chosen basis representation~\cite{Freysoldt2014}.

These limitations have motivated the application of \ac{WFT} to surface catalysis. Due to their computational cost and the challenges associated with periodic implementations, \ac{WFT} methods are typically combined with embedding approaches, in which a localized region of interest is treated at a higher level of theory, while the surrounding environment is described approximately. A range of embedding schemes exists, from Green’s function-based methods~\cite{Maier2005, Knizia2012} to more approximate frozen-density~\cite{Wesolowski1993, Schreder2024} or point-charge models~\cite{Barandiaran1988, Sushko2010}, each offering different trade-offs between accuracy and computational cost.

Even within an embedding framework, however, the number of explicitly treated electrons remains too large to treat exactly (with a \ac{FCI} description). This necessitates approximate \ac{WFT} methods, such as \ac{CC} or \ac{CAS}, where only a subset of the total correlation is treated explicitly~\cite{Roos1980, Bartlett2007}. The combination of embedding techniques with \ac{WFT} methods is therefore becoming increasingly common in solid state quantum chemistry~\cite{Knizia2012, Libisch2014, Larsson2022, Popov2023, Schreder2024, Lavroff2025, Kolodzeiski2025}.

In parallel, efforts to obtain \ac{FCI}-quality results for larger systems have driven interest in quantum computing approaches, including solid-state applications~\cite{Cao2023, Battaglia2024}, most notably within the \ac{VQE} framework~\cite{Peruzzo2014}. In principle, quantum algorithms offer a route to reduce the exponential scaling of \ac{FCI} to polynomial scaling~\cite{Aspuru-Guzik2005}. While current hardware remains far from achieving this goal for realistic systems, quantum algorithms designed as replacements for \ac{WFT} on classical computers have reached a level of maturity where they can be meaningfully tested on non-trivial chemical problems. Many systems of interest in surface catalysis exhibit strong multiconfigurational character and involve multiple closely lying electronic states along reaction coordinates, placing stringent demands on approximate methods.

The purpose of this work is to assess the performance of two commonly used quantum computing ansätze, \ac{fUCC}~\cite{Bartlett1989, Evangelista2019, Lee2019, Romero2019} and the adaptive, problem-tailored ansatz (ADAPT)~\cite{Grimsley2019, Tang2021}, in such a setting. As a reference, we employ \ac{SA} \ac{CASSCF} (SA-CASSCF), which provides an exact treatment of correlation within a chosen active space~\cite{Docken1972, Werner1981}. To isolate the performance of the wavefunction ansätze from orbital relaxation effects, we retain the SA-CASSCF orbitals and focus solely on the convergence of the corresponding CI problem. While multiple approaches exist for computing excited states with quantum ansätze (see Refs.~\citenum{Nakanishi2019, Parrish2019, Ollitrault2020,Zhang2022-cw}), we here restrict ourselves to \ac{SA} formulations and evaluate the performance of \ac{fUCC}\cite{Yalouz2022-xz,yalouz2021state} and ADAPT\cite{Grimsley2025-ie,trustmedude} as alternative \ac{CASCI} solvers within a fixed active space.

As a chemically motivated test system designed to probe multiconfigurational behavior in a surface-like environment, we consider NO adsorption and desorption on Rh-doped rutile TiO$_{2}$(110) (Rh@TiO$_{2}$(110)). This system has previously been studied using periodic \ac{DFT} under various conditions~\cite{Tang2019}. We select NO as the adsorbate because, unlike the species considered in Ref.~\citenum{Tang2019}, it is a radical. In the dissociated limit, the system can be viewed as two radical fragments: an unpaired electron on the formal Rh(IV) center and an NO$^{*}$ radical, forming an entangled state. In contrast, the adsorbed configuration resembles a closed-shell Rh(III)–NO$^{+}$ pair. This leads to significant changes in the electronic structure along the adsorption coordinate, with multiple closely lying electronic states and pronounced charge-transfer character. Consequently, the system provides a stringent test for methods requiring state averaging.

We emphasize that the primary objective of this work is methodological rather than chemical. The Rh@TiO$_{2}$–NO system is employed as a chemically motivated test case that captures key features relevant to state-averaged electronic structure problems, including multiconfigurational character, charge transfer, and state crossings. Accordingly, the focus is on assessing the performance of quantum computing ansätze within a controlled and internally consistent framework, rather than on providing a quantitatively converged description of surface chemistry.

\section{Computational details}

In the embedded cluster calculations, the Rh-doped TiO$_{2}$(110) surface was represented using \acp{AIMP}~\cite{Barandiaran1988} and point charges, generated with the SCEPIC program~\cite{Larsson2022}. A point-charge cutoff of 80 Å and an AIMP cutoff of 15 Å were applied around the Rh center, which substitutes a five-coordinated surface Ti site. The QM region consists of [RhO$_{5}$]$^{6-}$ and NO, embedded in 780 \acp{AIMP} and 105451 point charges. No geometry optimizations were performed (see the \ac{SI}). The rutile cell parameters were taken from the Materials Project (mp-2657)~\cite{Jain2013, MaterialsProject}.

The one-particle wavefunctions were expanded in the X2C-SVPAll basis set~\cite{PollakPatrik2017SCEB}, together with the scalar-relativistic X2C Hamiltonian and a finite-nucleus model~\cite{Visscher1997}. \acp{AIMP} were generated from the uncontracted primitives of the same basis set using a finite-nucleus model, ensuring consistency across all calculations.

SA-CASSCF calculations were performed with OpenMolcas~\cite{Fdez.Galvan2019, Aquilante2020}, using integrals from the SEWARD module. Layered SA-fUCCSD (i.e., SA-fUCCSD($n$)) and SA-ADAPT calculations were carried out with SlowQuant~\cite{SlowQuant}, employing one- and two-electron integrals generated with the Ädelsten library. Although restricted to single and double excitations, the SA-fUCCSD($n$) ansatz can approach the CASCI limit within a fixed orbital basis as the number of layers increases. In this work, we consider ansätze with six, eight, and ten layers (see Results and Discussion).

All calculations were restricted to singlet-spin states. In SA-CASSCF, this was achieved by expanding the wavefunction in \acp{CSF}. For SA-fUCCSD and SA-ADAPT, the reference state was taken to be the leading \ac{CSF} from the SA-CASSCF calculations, and only spin-adapted excitation operators were included in the operator pool\cite{Magoulas2025-wb,Kjellgren2025-pb}.
The state-resolution was resolved using the multistate contracted \ac{VQE} scheme\cite{Parrish2019}.
The implementation details for the SA-ADAPT variant in SlowQuant will be in a future publication\cite{trustmedude}. % This can just be resolved to an arXiv link when implementing review reports.

Cluster size convergence tests were performed using the PBE functional and compared against PBC-DFT calculations carried out with the CP2K package~\cite{Kuhne2020}, employing the geometrical response double-$\zeta$ basis set~\cite{Hutter2024ccgrb}, Goedecker–Teter–Hutter pseudopotentials~\cite{GTH1996}, and a plane-wave cutoff of 400 Ry. In addition to cluster size, basis set sensitivity was assessed using the cc-pVDZ(-PP)~\cite{Dunning1989, Balabanov2006, Peterson2007} and def2-SVP~\cite{Weigend2005, Andrae1990} basis sets, both employing effective core potentials for Rh.

The PBC calculations were performed on a $4\times2\times2$ supercell of the TiO$_{2}$(110) unit cell, with k-point sampling restricted to the $\Gamma$-point. The [110] direction was padded with four additional empty layers to introduce vacuum along the surface normal, resulting in a slab separation of approximately 26 Å. The PBC calculations used the same cell parameters as the embedded cluster calculations.

Due to differences in basis representations and boundary conditions, quantitative agreement between PBC-DFT and embedded cluster calculations is not expected; the comparison is therefore restricted to overall trends with increasing cluster size. As discussed in the \ac{SI}, the embedded cluster model is not fully converged with respect to size or polarization effects, but provides a consistent framework for assessing relative trends and method performance.

\section{Results and Discussion}

The following discussion serves a dual purpose: to characterize the electronic structure of the chosen model system and to define a consistent reference against which the performance of the quantum ansätze can be evaluated. We therefore begin with the classical SA-CASSCF results, which establish the active space and provide reference energies for the subsequent SA-fUCCSD and SA-ADAPT calculations.

All calculations consider NO adsorption and desorption on Rh-doped TiO$_{2}$(110), modeled using a minimal QM region consisting of [RhO$_{5}$]$^{6-}$ and NO. The adsorption coordinate is described along a simplified and largely frozen pathway, in which neither the surface geometry nor the internal NO coordinates are optimized. This choice isolates electronic structure effects and ensures consistency between methods (see \ac{SI} for detailed justification).

The resulting adsorption profile is not intended to represent a fully converged reaction pathway. Rather, it provides a controlled test case that captures the essential multiconfigurational character of the system while enabling a systematic assessment of the SA-fUCCSD and SA-ADAPT ansätze. At the same time, the model retains key qualitative features associated with catalytic environments, including transition metal chemistry, charge transfer, and state crossings, thereby allowing us to probe the performance of quantum algorithms in a chemically motivated setting.

\subsection{Classical SA-CASSCF results}

Since the remainder of our results are discussed relative to classical CASSCF references, and as SA-CASSCF orbitals are used for both SA-fUCCSD and SA-ADAPT calculations, we begin by outlining the selection of the active space and the included states.

The ``canonical'' active space for Rh–NO consists of the five Rh 4$d$ orbitals and the NO valence orbitals ($\sigma$, $\sigma^{*}$, $\pi$, $\pi^{*}$), corresponding to a (12,11) active space for [Rh–NO]$^{4+}$~\cite{Veryazov2011}. This space is well within the capabilities of classical CASSCF. To describe the crossover between a bound closed-shell state and a dissociated open-shell state, two states were initially included in a state-averaged CASSCF treatment.

However, analysis of the resulting wavefunctions revealed that the NO $\sigma$/$\sigma^{*}$ orbitals remain weakly correlated, with occupations close to (2.0, 0.0), and do not participate in the state crossing. Similarly, the Rh 4$d_z$ orbital remains essentially unoccupied and exhibits a tendency to rotate out of the active space. Furthermore, beyond a Rh–N distance of approximately 2.2 Å, the closed-shell state is no longer retained within a two-state SA-CASSCF description, indicating an insufficient state-averaging space.

Expanding the calculation to three states restores the closed-shell solution but does not alter the qualitative orbital analysis. Based on these observations, the NO $\sigma$/$\sigma^{*}$ orbitals and the Rh 4$d_z$ orbital were removed, yielding a reduced (10,8) active space. Such an active space is on the larger side for near-term, VQE-based quantum computing applications, but is at present still feasible on ideal statevector simulators.

\begin{figure}
    \centering
    \includegraphics[width=0.5\linewidth]{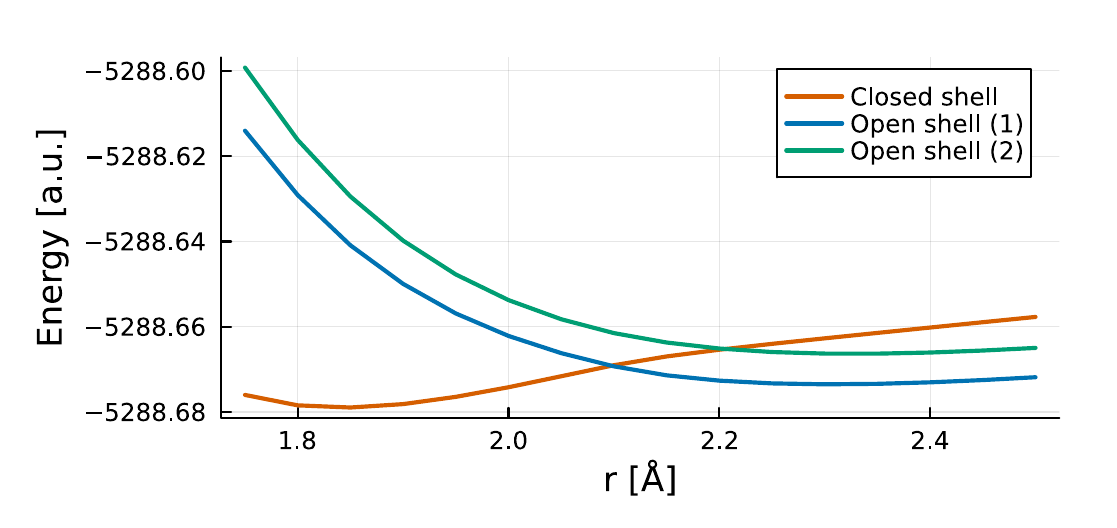}
    \caption{SA-CASSCF (10,8) adsorption profile for NO on Rh@TiO$_{2}$(110).}
    \label{fig:casscf_curves}
\end{figure}

Our final SA-CASSCF curves, based on the (10,8) active space and using three roots, are presented in Figure~\ref{fig:casscf_curves}. The CASSCF results suggest a bonded closed-shell solution at around 1.85 Å, with the first closed-shell/open-shell crossing at around 2.10 Å, and the second at around 2.20 Å.

Although we refer to this state as ``closed-shell'' for convenience, it already exhibits significant open-shell character at the equilibrium geometry. The total weight of the closed-shell CSF is only around 50 \% of the CASCI expansion, highlighting its strongly multiconfigurational nature. The leading closed-shell CSF corresponds to a charge-transfer configuration, in which the surface Rh(IV) is reduced to Rh(III) and the adsorbed NO is oxidized to NO$^{+}$. For brevity, we refer to this configuration as the ionic CSF in the following.

\begin{figure}[H]
\centering
\begin{subfigure}{0.5\linewidth}
    \centering
    \includegraphics[width=\linewidth]{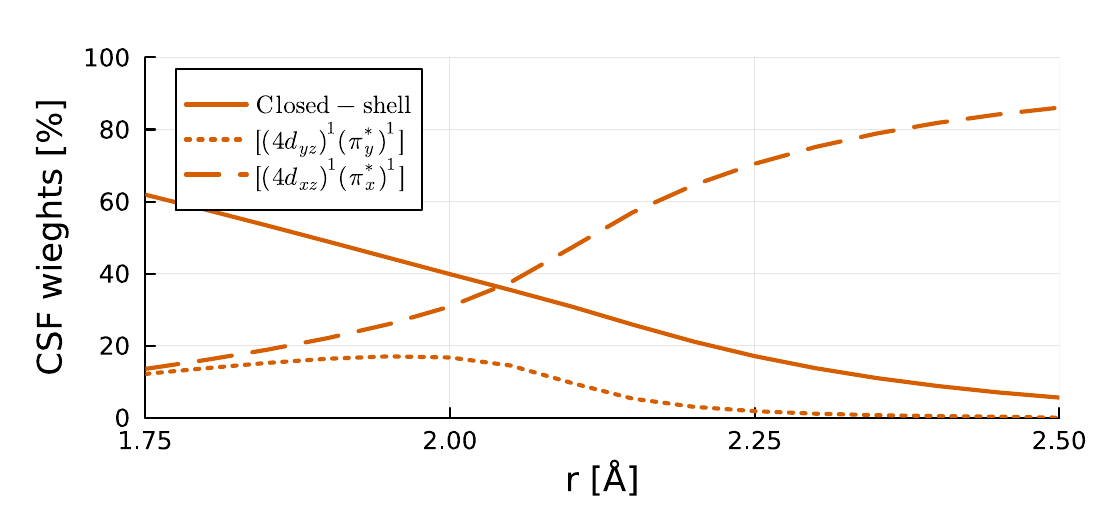}
    \caption{Closed-shell}
    \label{fig:state1_csf}
\end{subfigure}
\hfill
\begin{subfigure}{0.5\linewidth}
    \centering
    \includegraphics[width=\linewidth]{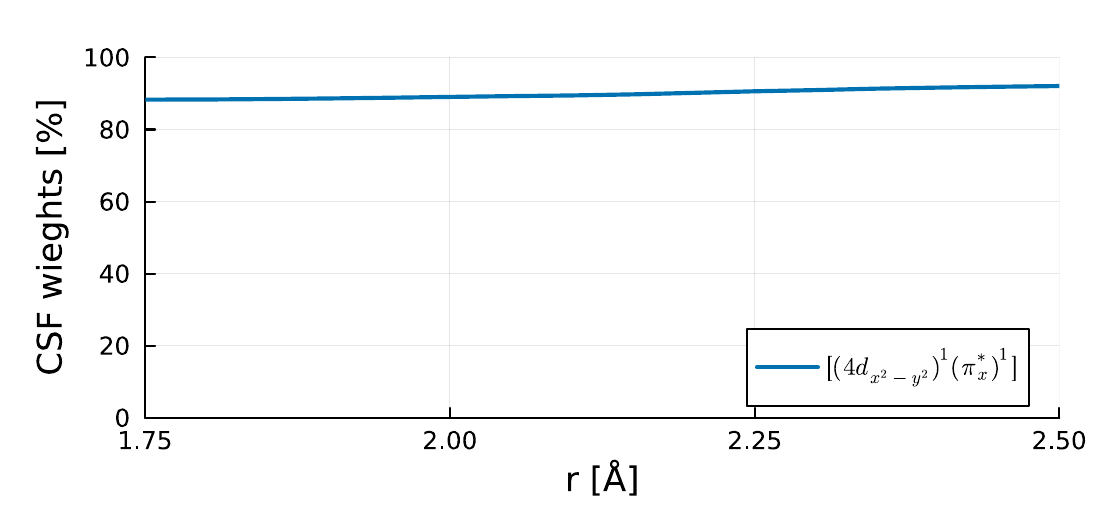}
    \caption{Open-shell (1)}
    \label{fig:state2_csf}
\end{subfigure}
\hfill
\begin{subfigure}{0.5\linewidth}
    \centering
    \includegraphics[width=\linewidth]{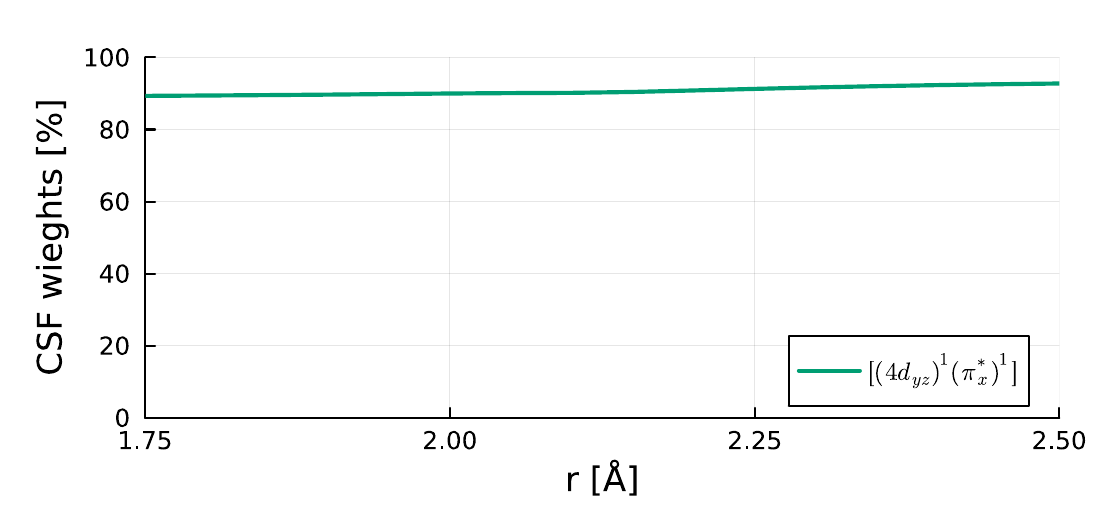}
    \caption{Open-shell (2)}
    \label{fig:state3_csf}
\end{subfigure}
\caption{\ac{CSF} weights along the adsorption/desorption pathway for the three states shown in Figure~\ref{fig:casscf_curves}. Open-shell \acp{CSF} are identified by their corresponding singly occupied orbitals. Only \acp{CSF} with weights exceeding 10\% in any configuration are included.}
\label{fig:csfs}
\end{figure}

In addition to the ionic CSF, excitations from the Rh 4$d_{xz}$ and 4$d_{yz}$ orbitals into the NO $\pi^{*}_{x}$ and $\pi^{*}_{y}$ orbitals contribute approximately 15\% and 20\%, respectively, to the CASCI expansion at 1.85 Å. As the Rh–N distance is increased, the weight of the Rh 4$d_{yz}$ $\rightarrow$ NO $\pi^{*}_{y}$ remains constant until around 2.1 Å (the first crossing point), after which this configuration decreases rapidly. In contrast, the ionic CSF decreases more smoothly, while the contribution from the Rh 4$d_{xz}$ $\rightarrow$ NO $\pi^{*}_{x}$ configuration increases steadily. The Cartesian orbital assignments reflect the geometry of the cluster and are not chosen to align with standard ligand field conventions. Beyond roughly 2.2 Å, it is therefore no longer accurate to describe this state as closed-shell, as it more closely corresponds to an open-shell singlet in which the two unpaired electrons occupy the Rh 4$d_{xz}$ and NO $\pi^{*}_{x}$ orbitals. However, given the smooth evolution of the CSF weights across the scan, we consider it reasonable to treat this as a single state throughout. The evolution of these weights, as well as those for the two open-shell states, are presented in Figure~\ref{fig:csfs}.

In contrast to the ``closed-shell'' solution, the two open-shell states are largely single-configurational in the CSF representation (although each corresponds to a spin-adapted linear combination of two Slater determinants) and remain consistent over the full scanning range. Their dominant contributions (approximately 88–90\% at 1.85 Å) correspond to excitations from the Rh 4$d_{x^{2}-y^{2}}$ and Rh 4$d_{yz}$ orbitals, respectively, into the NO $\pi^{*}_{x}$ orbital. Notably, at the largest Rh–N distances considered, all three states contain a single electron in the NO $\pi^{*}_{x}$ orbital. The primary difference in energy between these states therefore arises from the preferential occupation of the Rh 4$d_{x^{2}-y^{2}}$, 4$d_{yz}$, and 4$d_{xz}$ orbitals.

While the preferential occupation of the NO $\pi^{*}_{x}$ orbital over the NO $\pi^{*}_{y}$ orbital may at first appear surprising, it can be understood from the structure of the TiO$_{2}$(110) surface. In the idealized embedding employed here, the cluster exhibits $C_{2v}$ symmetry, which, as a non-degenerate point group, does not enforce any degeneracy between the $\pi^{*}$ orbitals.

\begin{figure}
    \centering
    \includegraphics[width=0.8\linewidth]{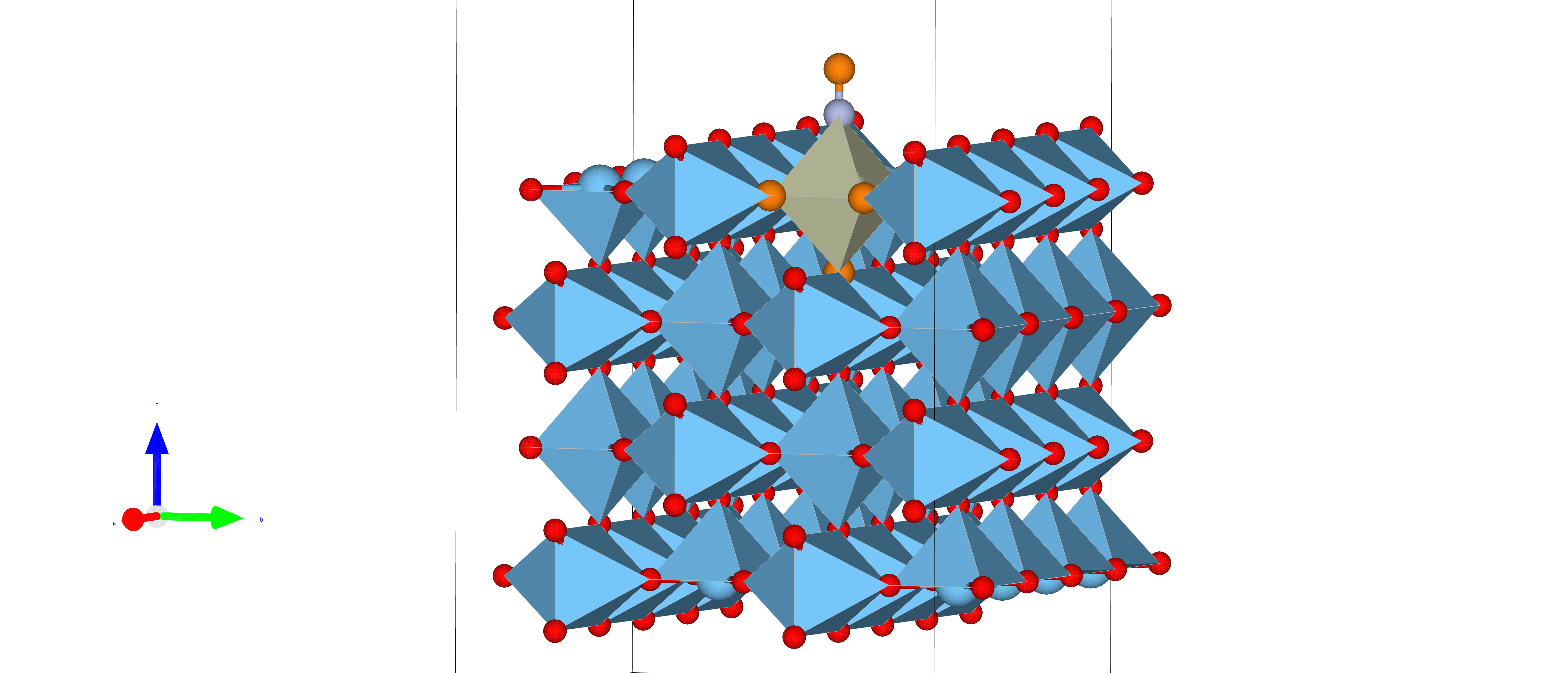}
    \caption{Visualization of NO adsorbed on Rh-doped TiO$_{2}$(110). The red, green, and blue arrows of the compass correspond to the [001], [$\bar{1}$10], and [110] crystallographic directions, respectively, and map onto the $x$-, $y$-, and $z$-axes in the cluster calculations. The blue and red octahedra correspond to regular TiO$_{6}$ units, whereas the bronze octahedron represents the doped site, to which the NO molecule is adsorbed.}
    \label{fig:surface}
\end{figure}

More importantly, the TiO$_{2}$(110) surface features rows of oxygen atoms running along the [001] direction, with periodic spacing in the [$\bar{1}$10] direction (see Figure~\ref{fig:surface}). In our coordinate system, the surface normal (z-axis) corresponds to the [110] direction, while the x- and y-axes align with the [001] and [$\bar{1}$10] directions, respectively. The $\pi^{*}_{x}$ and $\pi^{*}_{y}$ orbitals therefore correspond to orbitals oriented parallel and perpendicular to the surface rows of the TiO$_{2}$(110) facet. As a result, the NO $\pi^{*}_{y}$ orbital, whose lobes are oriented along the $y$-axis, points towards these surface features and experiences stronger electrostatic repulsion. In contrast, the $\pi^{*}_{x}$ orbital is oriented parallel to the surface rows and interacts less directly with the underlying oxygen density, leading to its relative stabilization.

Importantly, this system presents several features that make it a stringent test for quantum algorithms. The electronic structure is strongly multiconfigurational, with significant contributions from multiple configurations even at the equilibrium geometry. Furthermore, the relevant states exhibit substantial changes in character along the reaction coordinate, including charge-transfer and metal-to-ligand excitation contributions. Finally, the presence of multiple closely lying states necessitates a state-averaged description. Together, these aspects make this system particularly well-suited for assessing the performance of state-averaged quantum algorithms such as SA-fUCCSD and SA-ADAPT.

The resulting (10,8) active space corresponds to 16 qubits in a Jordan-Wigner mapping. While such system sizes are readily accessible to classical simulation, they already represent a significant challenge for current noisy quantum hardware, particularly in a state-averaged setting where circuit depth, measurement cost, and optimization complexity increase substantially. As such, this system occupies an intermediate regime: large enough to be non-trivial for quantum algorithms, yet still amenable to controlled studies on ideal simulators.

\subsection{Simulated SA-fUCCSD results}

We now turn to the SA-fUCCSD results. As noted above, we do not consider orbital-optimized SA-fUCCSD in this work. Instead, we retain the SA-CASSCF orbitals and restrict the optimization to the SA-fUCCSD wavefunction parameters alone, i.e., we focus solely on the state energies.

We consider this a physically motivated choice. In the limit of a sufficiently expressive SA-fUCCSD ansatz (i.e., with increasing layers, $n$), SA-fUCCSD($n$) converges towards the SA-CASCI solution within the given active space and fixed orbital basis. In this sense, the SA-CASSCF orbitals define the reference \ac{MO} basis, while the SA-fUCCSD ansatz serves as an approximate solver for the corresponding CASCI problem. Deviations from the SA-CASSCF energies can therefore be attributed primarily to ansatz incompleteness. In the limit of a complete ansatz, the use of CASSCF orbitals ensures recovery of the exact CASCI (and thus CASSCF) solution; however, for a truncated ansatz, the absence of orbital relaxation may introduce a secondary, albeit expectedly small, source of error.

Consequently, our focus is on assessing whether a tractable number of layers is sufficient to approach this limit, without introducing additional complexity from orbital optimization. We note that all error metrics reported in the following are defined relative to the SA-CASSCF reference within the fixed active space and orbital basis. As such, they reflect the ability of the ansätze to approximate the corresponding \ac{CASCI} problem, rather than the absolute accuracy of the underlying electronic structure model. We begin by considering a six-layer ansatz (810 spin-adapted excitation operators). At this depth, the SA-fUCCSD curves qualitatively resemble the SA-CASSCF curves: the equilibrium geometry is found at around 1.85 Å, and the two state-crossing points at around 2.10 Å and 2.20 Å, respectively.

While the qualitative agreement is good, there are some subtle but non-negligible differences. The overall \ac{MAD} is around 0.3 mE$_{h}$ for the closed-shell state and 0.2 mE$_{h}$ for the first open-shell state, indicating incomplete convergence. The largest deviations occur near the first crossing point, where the errors reach approximately 1.0 mE$_{h}$ and 0.6 mE$_{h}$ for the closed-shell and first open-shell states, respectively, suggesting that these regions are particularly challenging for the ansatz.

We note that there is substantial discussion in the literature on the sensitivity of fUCCSD, and other quantum ansätze, to the initial $\theta$ values~\cite{McClean2018, Grant2019, Grimsley2023}, where poor initialization can lead to convergence to local minima. In the results discussed so far, all $\theta$ values were initialized to zero. If, instead, we initialize the crossing point using the optimized parameters from the previous geometry (e.g., 2.05 Å), the errors for both states are reduced to approximately 0.15 mE$_{h}$.

\begin{figure}[H]
\centering
\begin{subfigure}{0.45\linewidth}
    \centering
    \includegraphics[width=\linewidth]{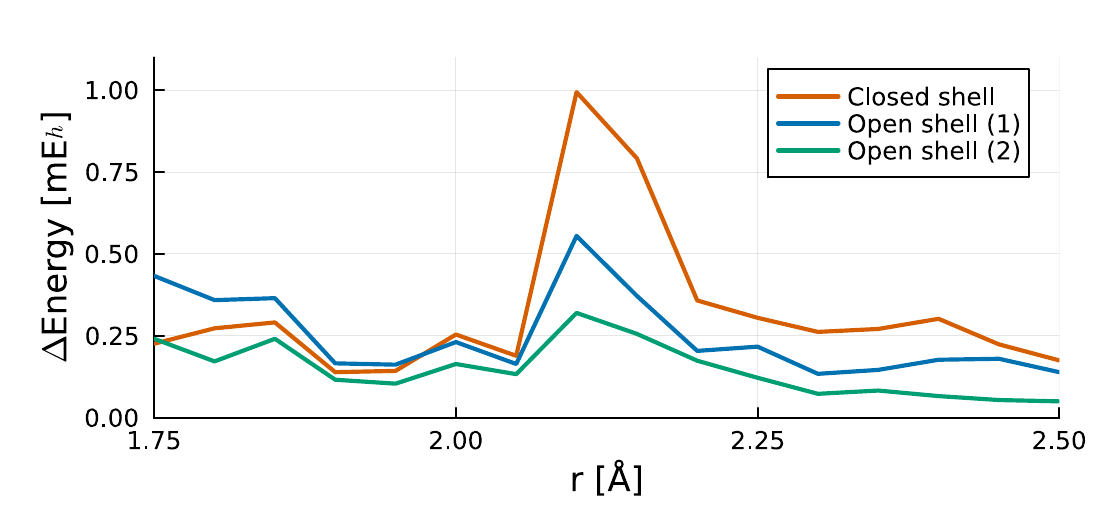}
    \caption{}
    \label{fig:uccsd_6layers_error}
\end{subfigure}
\begin{subfigure}{0.45\linewidth}
    \centering
    \includegraphics[width=\linewidth]{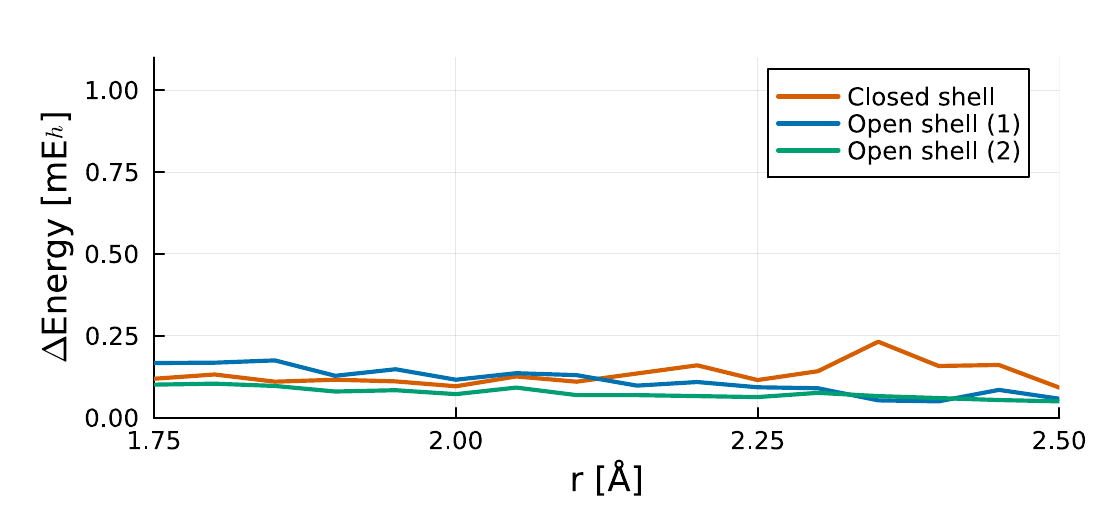}
    \caption{}
    \label{fig:uccsd_8layers_error}
\end{subfigure}
\hfill
\begin{subfigure}{0.45\linewidth}
    \centering
    \includegraphics[width=\linewidth]{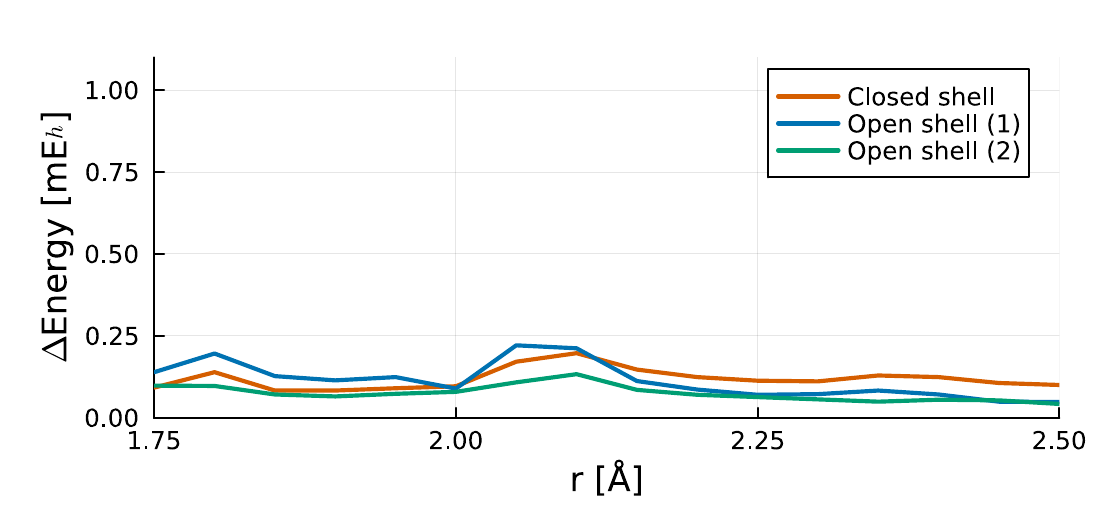}
    \caption{}
    \label{fig:uccsd_10layers_error}
\end{subfigure}
\begin{subfigure}{0.45\linewidth}
    \centering
    \includegraphics[width=\linewidth]{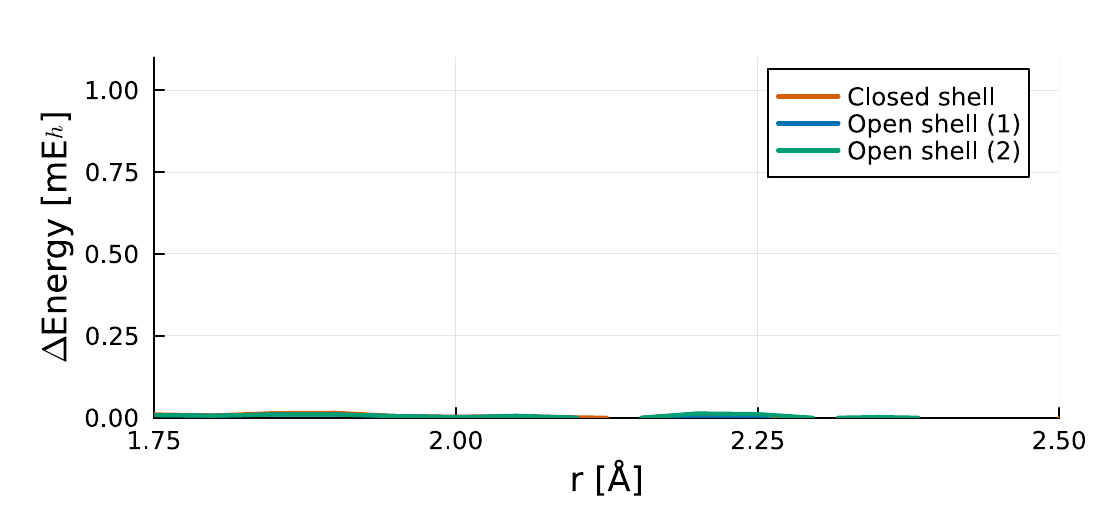}
    \caption{}
    \label{fig:adapt_grad5_error}
\end{subfigure}
\caption{Errors of SA-fUCCSD and SA-ADAPT relative to the SA-CASSCF reference curves shown in Figure~\ref{fig:casscf_curves}. Panels (a)–(c) correspond to SA-fUCCSD with six, eight, and ten layers, respectively, while panel (d) shows the SA-ADAPT results.}
\label{fig:errors}
\end{figure}

While this improves the results, the remaining error is still above our desired threshold. We therefore also considered increasing the number of layers to eight and ten, in both cases using zero initialization. At these depths, the sensitivity to geometry is reduced, and the \ac{MAD} decreases to 0.12 mE$_{h}$ and 0.11 mE$_{h}$, respectively. As can be seen in Figure~\ref{fig:errors}, the error profile shifts between six and eight layers, which is most likely due to the parameter initialization.

While this supports the use of deeper ansätze, the overall agreement of 0.11 mE$_{h}$ remains unsatisfactory, particularly given the large number of parameters (1350 for ten layers). Furthermore, the observed sensitivity to the initial $\theta$ values, especially in the vicinity of the state crossings, suggests that optimization becomes increasingly challenging as the ansatz grows in size. While previous work suggests that sampling a large number of random initial $\theta$ values may improve convergence~\cite{Boy2025}, we consider this approach computationally inefficient, as it would require either performing a large number of optimizations at each geometry or relying on parameter transfer between geometries without guaranteed robustness.

These observations indicate that simply increasing the depth of a fixed ansatz leads to diminishing returns, both in terms of accuracy and computational cost. We therefore turn to the ADAPT approach, which constructs the ansatz iteratively and may provide a more compact and robust representation of the wavefunction.

\subsection{Simulated SA-ADAPT results}

We now turn to the SA-ADAPT results. To remind the reader, in ADAPT the operators are drawn from the same pool as in fUCCSD; however, instead of adding operators in fixed layers, a single operator is added at each macroiteration. This operator is chosen as the one with the largest energy gradient, after which all parameters are re-optimized. While this approach may require many optimization cycles, it can, in principle, yield significantly more compact ansätze that converge toward the CASCI limit.

The ability of ADAPT to achieve improved accuracy with fewer operators has been discussed elsewhere~\cite{Grimsley2019}, but is also suggested by our SA-fUCCSD results. In particular, we observed that increasing the number of layers does not lead to uniform improvement, and in some cases, results obtained with eight layers outperform those obtained with ten layers, indicating limitations in fixed ansatz constructions.

In its naive form, ADAPT adds only a single operator per iteration. However, this introduces a degree of arbitrariness, as multiple operators often exhibit gradients of similar magnitude. Furthermore, ADAPT is prone to getting stuck in local minima, sometimes called ``gradient troughs''~\cite{Feniou2023,Grimsley2023}. Typically, a substantial number of iterations are required to break out of these troughs, leading to an over-parametrized ansatz.

We observe the same behavior in our system. Using the standard ADAPT procedure, the convergence slows significantly after approximately 100 iterations per geometry, with the energies remaining around 1 mE$_{h}$ above the SA-CASSCF reference. Beyond this point, the algorithm effectively stalls and fails to reach our desired convergence threshold of a maximum gradient norm below 10$^{-5}$ E$_{h}$.

To address this, we modified the ADAPT procedure by including all operators whose gradient norms are within 90\% of the maximum at each iteration. This reduces the arbitrariness of single-operator selection and accelerates convergence by incorporating multiple relevant directions simultaneously. The 90\% threshold was chosen based on preliminary testing, where it provided a reasonable balance between improving convergence and avoiding excessive growth of the ansatz.

\begin{figure}
    \centering
    \includegraphics[width=0.5\linewidth]{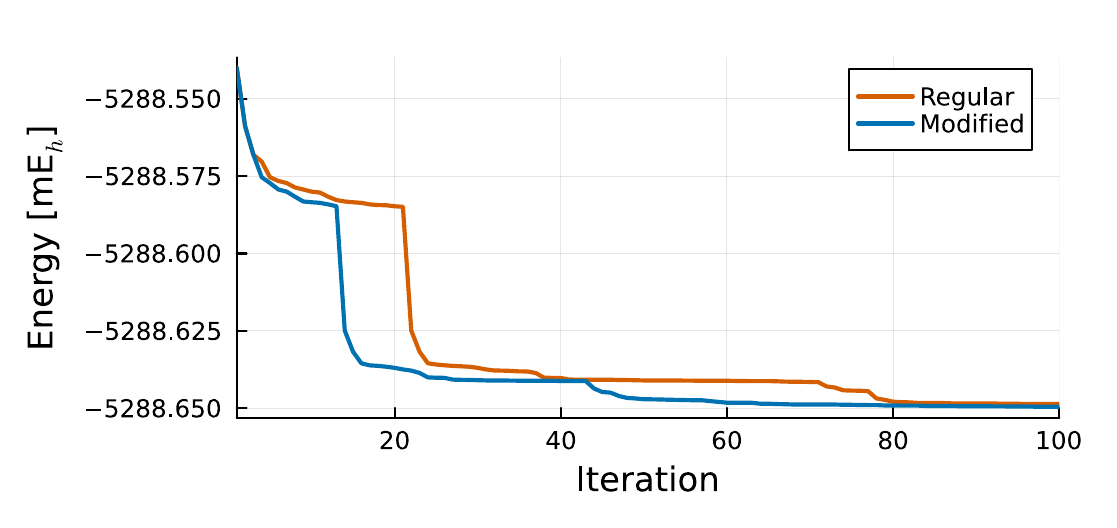}
    \caption{Convergence profiles at the equilibrium geometry with SA-ADAPT.}
    \label{fig:adapt_convergence}
\end{figure}

We focus here on the convergence behavior at the equilibrium geometry, and report the convergence behaviour in Figure~\ref{fig:adapt_convergence}; analogous results for other geometries are provided in the \ac{SI} (Figure S2). With this modified approach, the convergence improves noticeably: an accuracy of approximately 1 mE$_{h}$ is reached after around 60 iterations, compared to 80 iterations for the standard ADAPT procedure. At this stage, 101 operators have been included in the modified ADAPT. Beyond this point, both methods exhibit over-parametrization behavior, but the modified approach continues to converge and reaches the desired threshold after 212 iterations, having added a total of 386 operators, which was reached after roughly three days of calculations. In contrast, the standard ADAPT procedure did not reach this threshold even after 242 iterations (i.e., 242 operators), and we terminated the calculation after it had been running for over a week, as the convergence had effectively stalled.

Notably, this represents a substantial reduction compared to the SA-fUCCSD ansatz, where even the six-layer case involved 810 operators. Moreover, the modified ADAPT approach achieves agreement with the SA-CASSCF energies to within approximately 1 $\mu$E$_{h}$, corresponding to a reduction of the residual error within the fixed \ac{CASCI} problem by roughly two orders of magnitude. The errors are visualized in Figure~\ref{fig:errors}d, and the number of parameters for the various methods are presented in Table~\ref{tab:method_comparison}. At this level of accuracy, the remaining deviations are within numerical noise for total energies of approximately -5288 E$_{h}$. 

These results suggest that adaptive ansätze provide a significantly more efficient route to accurate state-averaged wavefunctions in this system. However, several challenges remain. The modified ADAPT procedure still exhibits over-parametrization behavior, indicating that more sophisticated operator selection strategies may be required. Alternative approaches, such as rotosolve-based optimization~\cite{Ostaszewski2021structure}, may further improve convergence. Additionally, the choice of convergence threshold remains somewhat arbitrary; relaxing the gradient norm threshold to 10$^{-4}$ E$_{h}$ leads to errors on the order of 10 $\mu$E$_{h}$. While such deviations are negligible for total energies, they may become more significant when considering derivative properties, and should therefore be chosen with care.

\begin{table}[H]
    \centering
    \begin{tabular}{c|ccc}
        \hline
         Method & Operators & MAD (mE$_{h}$) & Max Error (mE$_{h}$) \\
         \hline \hline
         SA-fUCCSD(6) & \phantom{0}810 & 0.32\phantom{0} & 0.79\phantom{0} \\
         SA-fUCCSD(8) & 1080 & 0.13\phantom{0} & 0.23\phantom{0} \\
         SA-fUCCSD(10) & 1350 & 0.12\phantom{0} & 0.22\phantom{0} \\
         SA-ADAPT & \phantom{00}289 & 0.004 & 0.015 \\
         \hline
    \end{tabular}
    \caption{Comparison of SA-fUCCSD and SA-ADAPT performance. SA-fUCCSD($n$) denotes an $n$-layer ansatz. The number of operators for SA-ADAPT corresponds to the average over all geometries (rounded to the nearest integer), with a minimum of 180 and a maximum of 420 operators. For reference, the SA-CASSCF expansion contains 1176 \acp{CSF}.}
    \label{tab:method_comparison}
\end{table}

\section{Conclusions}
We have assessed the performance of state-averaged fUCCSD and ADAPT ansätze for a chemically motivated, multiconfigurational surface model based on NO adsorption on Rh@TiO$_{2}$(110), using SA-CASSCF as a reference within a fixed active space.

SA-fUCCSD systematically reduces the deviation from the CASCI limit with increasing circuit depth. However, this improvement is slow and requires a large number of parameters, with the results exhibiting sensitivity to the choice of initial parameters. In contrast, the ADAPT ansatz achieves substantially lower residual errors with significantly fewer operators, demonstrating a more efficient representation of the wavefunction within the same active space.

The standard ADAPT protocol, however, was found to converge slowly for the present system, requiring a large number of iterations to reach the desired accuracy. A modified selection strategy, in which multiple operators are included per iteration based on their relative gradient magnitude, significantly accelerates convergence while retaining the compactness of the ansatz. These results suggest that the improved efficiency of adaptive ansätze arises not only from their reduced parameter count, but from their ability to prioritize the most relevant excitations early in the optimization. In contrast, layered fUCCSD constructions introduce a large number of redundant parameters, leading to slower convergence and increased sensitivity to initialization.

While demonstrated here for a single embedded cluster model, the observed trends are expected to extend to other multiconfigurational systems involving near-degenerate states and strong correlation. This work therefore provides a controlled benchmark for evaluating quantum ansätze in chemically motivated, multistate settings beyond minimal model systems.

%%%%%%%%%%%%%%%%%%%%%%%%%%%%%%%%%%%%%%%%%%%%%%%%%%%%%%%%%%%%%%%%%%%%%
%% The "Acknowledgement" section can be given in all manuscript
%% classes.  This should be given within the "acknowledgement"
%% environment, which will make the correct section or running title.
%%%%%%%%%%%%%%%%%%%%%%%%%%%%%%%%%%%%%%%%%%%%%%%%%%%%%%%%%%%%%%%%%%%%%
\begin{acknowledgement}
EDL and JK acknowledge financial support from the Danish e-Infrastructure Cooperation, case number 4317-00002B. JK further acknowledges financial support from the Novo Nordisk Foundation for the focused research project \textit{Hybrid Quantum Chemistry on Hybrid Quantum Computers} (HQC)$^2$, grant number NNFSA220080996.

\end{acknowledgement}

%%%%%%%%%%%%%%%%%%%%%%%%%%%%%%%%%%%%%%%%%%%%%%%%%%%%%%%%%%%%%%%%%%%%%
%% The same is true for Supporting Information, which should use the
%% suppinfo environment.
%%%%%%%%%%%%%%%%%%%%%%%%%%%%%%%%%%%%%%%%%%%%%%%%%%%%%%%%%%%%%%%%%%%%%
\begin{suppinfo}

The supporting information contains a discussion of embedded cluster size convergence relative to periodic density functional theory, as well as a justification for the use of a largely constrained adsorption geometry. In addition, convergence profiles for the SA-ADAPT calculations at all geometries are provided.

\end{suppinfo}

%%%%%%%%%%%%%%%%%%%%%%%%%%%%%%%%%%%%%%%%%%%%%%%%%%%%%%%%%%%%%%%%%%%%%
%% The appropriate \bibliography command should be placed here.
%% Notice that the class file automatically sets \bibliographystyle
%% and also names the section correctly.
%%%%%%%%%%%%%%%%%%%%%%%%%%%%%%%%%%%%%%%%%%%%%%%%%%%%%%%%%%%%%%%%%%%%%
\bibliography{main}

\end{document}